\documentclass[preprint,authoryear,3p,12pt]{elsarticle}

\usepackage{amssymb}
\usepackage{amsthm}

\usepackage{mathrsfs}
\usepackage{float}
\usepackage{eufrak}

\journal{}

\newtheorem{theorem}{Theorem}[section]

\newtheorem{define}[theorem]{Definition}
\newtheorem{example}[theorem]{Example}

\floatstyle{ruled}
\newfloat{algorithm}{!h}{loa}
\floatname{algorithm}{Algorithm}
\newcommand{\Input}[1]
  {\noindent\begin{tabular}{@{}p{1.8cm}@{}p{13.2cm}@{}}
   {\bf Input: }&#1 \end{tabular}}
\newcommand{\Output}[1]
  {\noindent\begin{tabular}{@{}p{1.8cm}@{}p{13.2cm}@{}}
   {\bf Output: }&#1 \end{tabular}}

\floatstyle{ruled}
\newfloat{function}{!h}{loa}
\floatname{function}{Function}

\floatstyle{ruled}
\newfloat{criterion}{!h}{loa}
\floatname{criterion}{Criterion}

\def\F{{\mathbb{F}}}
\def\N{{\mathbb{N}}}
\def\K{{\rm K}}

\def\lm{{\rm lm}}
\def\lpp{{\rm lpp}}
\def\lc{{\rm lc}}

\def\fb{{\bf f}}
\def\eb{{\bf e}}
\def\gb{{\bf g}}

\def\lcm{{\rm lcm}}

\def\deg{{\rm deg}}
\def\ind{{\rm index}}
\def\max{{\rm max}}
\def\min{{\rm min}}
\def\sp{{\rm spoly}}

\def\Q{{\mathbb{Q}}}
\def\gr{{Gr\"obner }}
\def\x{{x_1,\cdots,x_n}}
\def\prs{{\,<\,}}
\def\sus{{\,>\,}}

\def\s{{\rm{sign}}}
\def\p{{\rm{poly}}}

\def\lif{{\bf if \,}}
\def\lthen{{\bf then \,}}
\def\lelse{{\bf else \,}}
\def\lendif{{\bf end if\,}}

\def\lwhile{{\bf while \,}}

\def\lendwhile{{\bf end while\,}}
\def\ldo{{\bf do \,}}

\def\lend{{\bf end \,}}
\def\lreturn{{\bf return \,}}
\def\lla{{\longleftarrow}}

\def\lbegin{{\bf begin}}

\newcommand{\SPC}{\hspace*{15pt}}
\newcommand{\ignore}[1]{}

\begin{document}

\begin{frontmatter}



\title{The F5 Algorithm in Buchberger's Style}


\author{Yao Sun and Dingkang Wang\fnref{label1}}

\fntext[label1]{The authors are supported by NSFC 10971217,
10771206 60821002/F02.}

\address{Key Laboratory of Mathematics Mechanization, Academy of Mathematics and Systems Science, CAS, Beijing 100190,  China}

\ead{sunyao@amss.ac.cn, dwang@mmrc.iss.ac.cn}

\begin{abstract}
The famous F5 algorithm for computing \gr basis was presented by Faug\`ere in 2002. The original version of  F5 is given in
programming codes, so it is a bit difficult to understand. In this paper, the F5 algorithm is simplified as F5B in a Buchberger's
style such that it is easy to understand and implement. In order to describe F5B, we introduce F5-reduction, which keeps the
signature of labeled polynomials unchanged after reduction. The equivalence between F5 and F5B is also shown. At last, some
versions of the F5 algorithm are illustrated.
\end{abstract}


\end{frontmatter}



\section{Introduction} \label{sec-introduction}

Solving systems of polynomial equations is a basic problem in computer algebra, through which many practical problems can be
solved easily. Among all the methods for solving polynomial systems, the \gr basis method is one of the most efficient
approaches. After the concept of \gr basis is proposed in 1965 \citep{Buchberger65}, many algorithms have been presented for
computing \gr basis, including \citep{Lazard83, GebMol86, Gio91, Mora92, Fau99, Fau02}. Currently,   F5   is one of the most
efficient algorithms.

After the F5 algorithm is proposed, many researches have been done.
For example, Bardet et al. study the complexity of this algorithm in
\citep{Bardet03}. Faug\`ere and Ars use the F5 algorithm to attack
multivariable systems in \citep{Fau03}. Stegers revisits the F5
algorithm in his master thesis \citep{Stegers05}. Eder discusses the
two criteria of the F5 algorithm in \citep{Eder08} and proposes a
variation of the F5 algorithm \citep{Eder09}. Ars and Hashemi
present two variation of criteria in \citep{Ars09}. Recently, Gao et
al. give a new incremental algorithm in \citep{Gao09}. The current
authors discuss the F5 algorithm over boolean ring and present a
branch F5 algorithm \citep{SunWang09a, SunWang09b}. We also give a
complete proof for the correctness of the F5 or F5-like algorithm in
\citep{SunWang10}.

However, since the original F5 algorithm is reported in programming codes, it seems a bit difficult to understand this famous
algorithm very well. In this paper, we revisit the F5 algorithm from our perspective and we simplify F5 to F5B in Buchberger's
style,    which is equivalent to the original F5 algorithm.  We also discuss some versions of the F5 algorithm, such as {\it
Incremental} F5 algorithm in \citep{Fau02}, {\it Non-incremental} F5 algorithm reported by Faug\`ere in INSCRYPT 2008 and {\it
Matrix} F5 algorithm mentioned in \citep{Bardet03}.

This paper is organized as follows. The main idea of the F5 algorithm is illustrated in Section 3 after some basic notions given
in Section 2. In Section 4,  F5B  is proposed  and   the equivalence of  F5 and F5B is shown as well. Some versions of the F5
algorithm are discussed in Section 5. At last, some conclusions are presented in Section 6.

\section{Basic Notations}

Let $\K$ be a field and $\K[X]=\K[\x]$ a polynomial ring with
coefficients in $\K$. Let $\N$ be the set of non-negative integers
and $PP(X)$ the set of power products of $\{x_1,\cdots,x_n\}$,
i.e. $PP(X):=\{x^\alpha  \mid x^\alpha=x_1^{\alpha_1}\cdots
x_n^{\alpha_n}, \alpha_i \in \N, i=1,\cdots,n\}$.

Let $\prec$ be an admissible order defined over $PP(X)$. Given
$t=x^{\alpha} \in PP(X)$, the degree of $t$ is defined as
$\deg(t):= |{\alpha}| =\sum_{i=1}^n\alpha_i$. For a polynomial
$0\not=f\in \K[x_1,\cdots,x_n]$, we have $f=\sum
c_{\alpha}x^{\alpha}$. The degree of $f$ is defined as
$\deg(f):=\max\{| \alpha |, c_{\alpha}\not=0\}$ and the leading
power product of $f$ is $\lpp(f):=\max_\prec\{x^{\alpha},
c_{\alpha}\not=0\}$. If $\lpp(f)=x^{\alpha}$, then the leading
coefficient and leading monomial of $f$ are defined as
$\lc(f):=c_{\alpha}$ and $\lm(f):=c_{\alpha}x^{\alpha}$
respectively.

\section{Revisit the F5 Algorithm}

In brief, the major contribution of the F5 algorithm is presenting two new criteria:  Syzygy Criterion and  Rewritten Criterion.
Syzygy Criterion is  also called F5 Criterion in some papers.
   Almost all useless computations (redundant   S-polynomials) can be removed
by these two  criteria. Both of the criteria are built on the concept of signature and a special reduction procedure.

In this section, we first illustrate the main idea of Syzygy Criterion and Rewritten Criterion by some simple examples, and then
introduce the definition of signature. The special reduction is presented in next section. For more details about the F5
algorithm, please see \citep{Fau02}.

\subsection{About Syzygy Criterion} \label{subsec-aboutsyzygy}

Of the two   criteria, Syzygy Criterion is more important and creative. Let us see a simple example.

\ignore{
\begin{example} \label{exa-buchbergersyzygy}
Consider a system in $\Q[x, y]$ with the Graded Reverse Lex Order with $x\succ y$:
$$\left\{\begin{array}{l} f_1=x^2-y, \\
f_2=y^2-x.
\end{array}\right. $$
\end{example}
The S-polynomial $\sp(f_1, f_2)=y^2f_1-x^2f_2$ is not necessary to
reduce according to the Buchberger's criterion \citep{Buchberger79},
since the leading power products $\lpp(f_1)=x^2$ and $\lpp(f_2)=y^2$
are co-prime.

However, this fact can also be explained in a slightly different way. For  a   2-tuple vector $(f_1, f_2)$ in $\Q[x, y]^2$, there
exists a trivial syzygy $(f_2, -f_1)$, i.e.   $f_2f_1-f_1f_2=0$  or
$$\lm(f_2)f_1-\lm(f_1)f_2=(\lm(f_2)-f_2)f_1-(\lm(f_1)-f_1)f_2,$$ specifically,
$$\sp(f_1, f_2)=y^2f_1-x^2f_2=xf_1-yf_2,$$ which indicates the S-polynomial
$\sp(f_1, f_2)$ reduces to $0$ by $\{f_1, f_2\}$. The highlight of this explanation is that {\it it is the syzygy of $(f_1, f_2)$
that leads to   $\sp(f_1, f_2)$ reducing to $0$.}

In fact, the syzygy of $(f_1, f_2)$ can also result in some S-polynomials reducing to $0$ even if $\lpp(f_1)$ and $\lpp(f_2)$ are
not co-prime.

Now, let's take a look on the following example:}
\begin{example} \label{exa-syzygy}
Consider a system in $\Q[x, y, z]$ with the Graded Reverse Lex Order with $x\succ y\succ z$,
$$\left\{\begin{array}{l} f_1=x^2+y, \\
f_2=xy-z.
\end{array}\right. $$
\end{example}

The S-polynomial $\sp(f_1, f_2)=yf_1-xf_2=y^2+xz$, which is not reducible by $\{f_1, f_2\}$. We denote $f_3=\sp(f_1, f_2) $. Now,
consider the S-polynomial of $f_3$ and $f_2$: $$\sp(f_3, f_2)=xf_3-yf_2=x^2z+yz=zf_1,$$ the leading monomial of $f_2$ and $f_3$
are not co-prime, but   $\sp(f_3, f_2)$ reduces to $0$ by $\{f_1, f_2\}$.

{\bf Why? What makes this S-polynomial $\sp(f_3, f_2)$ reduce to $0$?}

If we dig it deeper, we will see that the reason still comes from
the syzygy $(f_2, -f_1)$ of the 2-tuple vector $(f_1, f_2)$. Since
$f_3=yf_1-xf_2$ and $f_2f_1-f_1f_2=0$, then we have
$$\sp(f_2,f_3)=xf_3-yf_2=xyf_1-(x^2+y)f_2=\lm(f_2)f_1-f_1f_2=(\lm(f_2)-f_2)f_1.$$
Thus, it is natural that the S-polynomial $\sp(f_3, f_2)$ reduces to $0$ by $\{f_1\}$.

Theoretically, in order to speed up the algorithm for computing \gr basis, we have to avoid computing the S-polynomials of the
above kind. Now, the  question is: {\bf how to detect them?} As the S-polynomial $\sp(f_3, f_2)$ reduces to $0$ due to the syzygy
$(f_2, -f_1)$ of the vector $(f_1, f_2)$, a natural idea is to {\em connect} the polynomials $f_3$ and $f_1$.

The relation $f_3=yf_1-xf_2$ means $f_3$ comes from $yf_1$, so we can append     a {\em signature} $y\eb_1$ to $f_3$ to reflect
this fact. Similarly, we also can append a signature $xy\eb_1$ to  $xf_3$ , which shows   $xf_3$ comes from $xyf_1=\lm(f_2)f_1$.
Now from the signature $xy\eb_1=\lm(f_2)\eb_1$, we are able to understand {\em why} the S-polynomial $\sp(f_3, f_2)$ can be
reduced to $0$.


In more general cases, for  the polynomial system $\{f_1, \cdots, f_m\}\subset \K[X]$, the syzygies of the $m$-tuple vector
$(f_1, \cdots, f_m)\in (\K[X])^m$ also result in many S-polynomials reducing to $0$. So the {\em Syzygy Criterion is a criterion
that detects useless S-polynomials by using syzygies of $(f_1, \cdots, f_m)$}. In the above simple example, the {\em signature}
of polynomial provides an useful information to detect this kind of unnecessary S-polynomials. Next, we give a mathematical
explanation of signatures.

\subsection{Signatures and Labeled Polynomials} \label{subsec-signature}

Consider a polynomial system $\{f_1, \cdots, f_m\}\subset \K[X]$ and denote $(f_1, \cdots, f_m)$ a polynomial $m$-tuple in
$(\K[X])^m$. We call the $f_i$'s initial polynomials of the ideal $\langle f_1, \cdots, f_m\rangle$, since they are initial generators of ideal $\langle f_1, \cdots, f_m
\rangle\subset \K[X]$.

Let $\eb_i$ be the canonical $i$-th unit vector
in the free $\K[X]$-module $(\K[X])^m$, i.e. the $i$-th element of $\eb_i$ is $1$, while the others are $0$. Consider the homomorphism
map $\sigma$ over the free $\K[X]$-module $(\K[X])^m$:
$$\sigma: (\K[X])^m \longrightarrow \K[X],$$
$$(g_1, \cdots, g_m) \longmapsto g_1f_1+\cdots+g_mf_m.$$ Then $\sigma(\eb_i)=f_i$. More
generally, if $\gb=g_1\eb_1+\cdots+ g_m\eb_m$, where $g_i\in \K[X]$
for $1\le i\le m$, then $\sigma(\gb)=g_1f_1+\cdots+g_mf_m$.

The admissible order $\prec$ on $PP(X)$ extends to the free module
$(\K[X])^m$ naturally in a POT (position over term) fashion:
$$x^\alpha\eb_i \prs x^\beta\eb_j (\mbox{ or } x^\beta\eb_j \sus x^\alpha\eb_i)\ \ \mbox{  iff  }
\left\{\begin{array}{l} i > j, \\ \mbox{ or } \\ i = j \mbox{ and
} x^\alpha \prec x^\beta.
\end{array}\right.
$$ Thus we have $\eb_m \prs \eb_{m-1} \prs \cdots \prs
\eb_1$ directly.

With the admissible order $\prs$ on $(\K[X])^m$, we can define the leading power product, leading coefficient and leading
monomial of a $m$-tuple vector $\gb\in (\K[X])^m$. For example, let $\gb=(2x^2+y^2, 3xy)\in (\Q[x, y])^2$ or equivalently
$\gb=(2x^2+y^2)\eb_1+3xy\eb_2$ and $\prec$ the Lex order on $PP(x, y)$ ($x \succ y$). Then we have $\lpp(\gb)=x^2\eb_1$,
$\lc(\gb)=2$ and $\lm(\gb)=2x^2\eb_1$.




The following is the definition of labeled polynomial and its signature. The idea of labeled polynomials draws lessons from
\citep{Gao09}.

\begin{define}[labeled polynomial]
Let $g\in \langle f_1, \cdots, f_m\rangle$ be a polynomial and $\gb\in (\K[X])^m$ an $m$-tuple vector such that $\sigma(\gb)=g$.
Then we call ${\mathcal G}=(\gb, g)$   a labeled polynomial.


For a labeled polynomial ${\mathcal G}$, we define
\begin{enumerate}
\item the signature $\s({\mathcal G}):=\lpp(\gb)$
\item the polynomial part $\p({\mathcal G}):=g$
\item the leading power product  $\lpp({\mathcal G}):=\lpp(g)$
\item  the leading monomial $\lm({\mathcal G}):=\lm(g)$
\end{enumerate}
\end{define}




 Suppose ${\mathcal F}=(\fb, f), {\mathcal G}=(\gb,
g) $ are labeled polynomials and  $cx^\alpha$ is a non-zero monomial, we define scaler multiplication and addition for labeled
polynomials as following.

\begin{itemize}
\item   $cx^\alpha   \cdot {\mathcal F}=cx^\alpha{\mathcal F} =(cx^\alpha \fb, cx^\alpha f)$.
\item   ${\mathcal F}+{\mathcal G}=(\fb+\gb, f+g)$.
\end{itemize}

\begin{define}[critical pair, S-polynomial]
 For labeled polynomials ${\mathcal F}, {\mathcal G}$, we say $[{\mathcal F}, {\mathcal G}]:=(u, {\mathcal F}, v, {\mathcal G})$ is a {\em critical pair} of ${\mathcal F}$ and
${\mathcal G}$ if $u, v$ are monomials   such that $u\lm({\mathcal F})=v\lm({\mathcal G})=\lcm(\lpp({\mathcal F}), \lpp({\mathcal
G}))$, and the corresponding {\em S-polynomial} of $[{\mathcal F}, {\mathcal G}]$ is denoted by $\sp({\mathcal F}, {\mathcal
G})=u{\mathcal F}-v{\mathcal G}$.
\end{define}

\subsection{Syzygy Criterion}

We are now able to describe the Syzygy Criterion in mathematical
words. We begin by the following definition.

\begin{define}[Divisible]
Let ${\mathcal F}=(\fb, f)$ be a labeled polynomial with
$\lpp(\fb)=x^\alpha\eb_i$, $cx^\gamma$ a non-zero monomial  and $B
$ a set of labeled polynomials. The labeled polynomial
$cx^\gamma{\mathcal F} $ is said to be divisible by $B$, if there
exists a labeled polynomial ${\mathcal G}=(\gb, g)\in B$ with
$\lpp(\gb)=x^\beta\eb_j$ such that

\begin{enumerate}

\item $\lpp(g) \mid x^{\gamma+\alpha}$, and

\item $\eb_i \sus \eb_j$, i.e. $i<j$.
\end{enumerate}
\end{define}

Then the Syzygy Criterion is described as follows.

\begin{criterion} \caption{\bf --- Syzygy Criterion}
  \smallskip
  \medskip

  \noindent
Let $[{\mathcal F}, {\mathcal G}]:=(u, {\mathcal F}, v, {\mathcal G})$ be a critical pair and $B $ a set of labeled polynomials.
If either $u{\mathcal F}$ or $v{\mathcal G}$ is {\bf divisible} by $B$, then the critical pair $[{\mathcal F}, {\mathcal G}]$
meets the Syzygy Criterion.
  \medskip
\end{criterion}

Syzygy criterion is also called F5 criterion in some papers. In  F5, if a critical pair meets the Syzygy Criterion, then the
corresponding S-polynomial is redundant.

 Let us review the two
simple examples from Subsection \ref{subsec-aboutsyzygy} in the view of Syzygy Criterion.

\ignore{
In Example \ref{exa-buchbergersyzygy}, Let ${\mathcal F}_1=(\eb_1, f_1)$ and ${\mathcal F}_2=(\eb_2, f_2)$. For the critical pair
$[{\mathcal F}_1, {\mathcal F}_2]=(y^2, {\mathcal F}_1, x^2, {\mathcal F}_2)$, it is easy to check $y^2{\mathcal F}_1$ is
divisible by $\{{\mathcal F}_2\}$. the critical pair $[{\mathcal F}_1, {\mathcal F}_2]$ meets the Syzygy Criterion, Therefore,
its S-polynomial is  redundant. This is correct, since its S-polynomial reduces to $0$ in Example \ref{exa-buchbergersyzygy}.}

In Example \ref{exa-syzygy}, Let ${\mathcal F}_1=(\eb_1, f_1)$ and ${\mathcal F}_2=(\eb_2, f_2)$. Let
  ${\mathcal F}_3=\sp ({\mathcal F}_1 , {\mathcal F}_2 )=y{\mathcal F}_1-x{\mathcal F}_2=(y\eb_1-x\eb_2, y^2+xz)$.

Next, consider the critical pair $[{\mathcal F}_3, {\mathcal F}_2]=(x, {\mathcal F}_3, y, {\mathcal F}_2)$. However, $x{\mathcal
F}_3=(xy\eb_1-x^2\eb_2, x(y^2+xz))$ is divisible by $\{ {\mathcal F}_2\}$, so the critical pair $[{\mathcal F}_3, {\mathcal
F}_2]$ is also redundant by Syzygy Criterion. And $\sp({\mathcal F}_3, {\mathcal F}_2)$ does reduce to $0$ in Example
\ref{exa-syzygy}, which verifies the Syzygy Criterion.

\subsection{Rewritten Criterion}

We next describe the Rewritten Criterion. In fact, Rewritten
Criterion is more like a programming technique, which reflects
Buchberger's idea: {\em try to reuse as much as possible the
previous computations.} Let us see the following example.

\begin{example} \label{exa-rew}
Consider a system in $\Q[x, y]$ with the Graded Reverse Lex Order with $x\succ y$:
$$\left\{\begin{array}{l} f_1=x^2+xy, \\
f_2=x^2+y.
\end{array}\right. $$
\end{example}

The S-polynomial $\sp(f_1, f_2)=f_1-f_2=xy-y$ is not reducible by
$\{f_1, f_2\}$. So denote $f_3=xy-y$. Next we need to consider the
S-polynomial $\sp(f_1, f_3)=yf_1-xf_3$ as well as the S-polynomial
$\sp(f_2, f_3)=yf_2-xf_3$. However, with a further study, we will
find that the S-polynomial $\sp(f_1, f_3)$ is not necessary to
reduce due to the following observation:
$$\sp(f_1, f_3)=yf_1-xf_3 = y(f_1-f_2)-(yf_2-xf_3)= y\, \sp(f_1, f_2)-\sp(f_2, f_3).$$ That is, if the
S-polynomials $\sp(f_1, f_2)$ and $\sp(f_2, f_3)$ have been reduced
(or scheduled to reduce), then reducing the S-polynomial $\sp(f_1,
f_3)$ becomes a duplicated job.\footnote{Rigorous proof needs the
knowledge of $t$-representation. For more details, please see the
authors' another paper \citep{SunWang10}.}

Thus, the {\em Rewritten Criterion is a criterion that detects
duplicated reductions of polynomials.} The meaning of Rewritten
Criterion is much clearer in a variation of the F5 algorithm that
uses matrix reduction, which will be discussed in Section
\ref{sec-presentversion}. Now we give the mathematical definition of
Rewritten Criterion.

\begin{define}[Rewritable]
Let ${\mathcal F}=(\fb, f) $ be a labeled polynomial, $cx^\gamma$ a non-zero monomial in $X$ and $B $ a set of labeled
polynomials. The labeled polynomial $cx^\gamma{\mathcal F}=(cx^{\gamma}\fb, cx^\gamma f)$ is said to be rewritable by $B$, if
there exists a labeled polynomial ${\mathcal G}=(\gb, g)\in B$, such that:
\begin{enumerate}

\item   $\s({\mathcal G})\mid \s(cx^\gamma{\mathcal F})$, i.e. $\lpp(\gb) \mid \lpp(cx^{\gamma}\fb)$, and

\item labeled polynomial ${\mathcal G}$ is {\bf generated later}
than ${\mathcal F}$.

\end{enumerate}
\end{define}

The Rewritten Criterion is given as follows.

\begin{criterion} \caption{\bf --- Rewritten Criterion}
  \smallskip
  \medskip

  \noindent
Let $[{\mathcal F}, {\mathcal G}]:=(u, {\mathcal F}, v, {\mathcal G})$ be a critical pair where $u, v$ are monomials in $X$ such
that $u\lm({\mathcal F})=v\lm({\mathcal G})=\lcm(\lpp({\mathcal F}), \lpp({\mathcal G}))$, and $B $ a set of labeled polynomials.
If either $u{\mathcal F}$ or $v{\mathcal G}$ is {\bf rewritable} by $B$, then the critical pair $[{\mathcal F}, {\mathcal G}]$
meets the Rewritten Criterion.
  \medskip
\end{criterion}

If a critical pair meets the Rewritten Criterion in the F5 algorithm, then it is not necessary to reduce its S-polynomial. Now
let us explain the fact in Example \ref{exa-rew}. The system is $f_1=x^2+xy, f_2=x^2+y\in \Q[x, y]$. Labeled polynomials
${\mathcal F}_1=(\eb_1, f_1)$ and ${\mathcal F}_2=(\eb_2, f_2)$ correspond to $f_1$ and $f_2$ respectively. The S-polynomial of
${\mathcal F}_1$ and ${\mathcal F}_2$ is ${\mathcal F}_3={\mathcal F}_1-{\mathcal F}_2=(\eb_1-\eb_2, xy-y)$.

Next, let us see the critical pair $[{\mathcal F}_1, {\mathcal F}_3]=(y, {\mathcal F}_1, x, {\mathcal F}_3)$. Clearly, labeled
polynomial ${\mathcal F}_3$ is {\em generated later} than ${\mathcal F}_1$, so it is easy to know $y{\mathcal F}_1$ is rewritable
by $\{{\mathcal F}_3\}$. Thus, critical pair $[{\mathcal F}_1, {\mathcal F}_3]$ meets the Rewritten Criterion, and hence, it can
be removed.

As discussed in this section, both Syzygy Criterion and Rewritten Criterion build on the concept of signature. However, to ensure
both criteria correct during the computations, a special reduction procedure is necessary, which is detailed in the next section.

\section{Simplify the F5 algorithm to a Buchberger's Style} \label{sec-F5B}

In order to simplify the F5 algorithm to a Buchberger's style, the most important step is to rewrite the reduction procedure. The
original reduction in the F5 algorithm is described by codes and may return several reduction results each time. In this section,
a simplified version of reduction (F5-reduction) is proposed in the first subsection. The F5 algorithm in Buchberger's style (F5B
algorithm) is described in the second subsection. We will discuss the equivalence of the F5 and F5B algorithms in the last
subsection.

\subsection{F5-Reduction}  \label{subsec-reduce}

The signatures of the labeled polynomials are used to detect   useless critical pairs by the criteria, but this is not sufficient
to ensure the correctness of the F5 algorithm. Only under F5-reduction, which is a special kind of reduction process, the
critical pairs detected by the criteria are really useless. The same is true for other F5-like algorithms.

Let us start with the definition of F5-reduction, which is a revised
version of the {\em TopReduction} function in \citep{Fau02}.

\begin{define}[F5-reduction] \label{df-reduce}
Let ${\mathcal F}=(\fb, f)  $ be a labeled polynomial and $B $ a set of labeled polynomials. The labeled polynomial ${\mathcal
F}$ is F5-reducible by $B$, if there exists ${\mathcal G}=(\gb, g)\in B$ such that: \footnote{Deleting the conditions 3 and 4
does not affect the correctness of algorithm, but leads to redundant computations/reductions.}
\begin{enumerate}

\item $\lpp(g) \mid \lpp(f)$, denote $x^\gamma=\lpp(f)/\lpp(g)$
and $c=\lc(f)/\lc(g)$,

\item $\s({\mathcal F})\sus \s(cx^\gamma{\mathcal G})$, i.e.
$\lpp(\fb) \sus \lpp(cx^{\gamma}\gb)$,

\item $cx^\gamma{\mathcal G}$ is {\bf not} divisible by $B$, and

\item $cx^\gamma{\mathcal G}$ is {\bf not} rewritable by $B$.
\end{enumerate}
If ${\mathcal F}$ is F5-reducible by $B$, let ${\mathcal
F}'={\mathcal F}-cx^\gamma{\mathcal G}$. Then this procedure:
${\mathcal F}\Longrightarrow_B {\mathcal F}'$ is called one step
F5-reduction. If ${\mathcal F}'$ is still F5-reducible by $B$, then
repeat this step until ${\mathcal F}'$ is not F5-reducible by $B$.
Suppose ${\mathcal F}^*$ is the final result that is not
F5-reducible by $B$. We say ${\mathcal F}$ F5-reduces to ${\mathcal
F}^*$ by $B$, and denote it as ${\mathcal F}\Longrightarrow_B^*
{\mathcal F}^*$.

\end{define}

The key of F5-reduction is the condition $\s({\mathcal F})\sus
\s(cx^\gamma{\mathcal G})$, i.e. $\lpp(\fb) \sus
\lpp(cx^{\gamma}\gb)$, which makes F5-reduction much different from
other general reductions. The major function of this condition is to
preserve the signature of ${\mathcal F}$ during reductions. Thus a
direct result is that, if labeled polynomial ${\mathcal F}$
F5-reduces to ${\mathcal F}^*$ by $B$ (i.e. ${\mathcal
F}\Longrightarrow_B^* {\mathcal F}^*$), then the signatures of
${\mathcal F}$ and ${\mathcal F}^*$ are identical, i.e.
$$\s({\mathcal F})=\s({\mathcal F}^*).$$ This property plays a
crucial role in the proof for the correctness of the F5B algorithm.
For more details, please see \citep{SunWang10}.

\subsection{The F5 algorithm in Buchberger's style} \label{subsec-buchstyle}

With the definitions of Syzygy Criterion, Rewritten Criterion and
F5-reduction, we can rewrite the F5 algorithm in Buchberger's style
(F5B algorithm).

\begin{algorithm} \caption{\bf --- The F5 algorithm in Buchberger's style (F5B algorithm)} \label{algorithm1}
  \smallskip
  \Input{a polynomial $m$-tuple: $(f_1,\cdots,f_m)\in K[X]^m$; and an admissible order $\prec$.}\\
  \Output{The \gr basis of the ideal $\langle f_1,\cdots,f_m\rangle \subset K[X]$.}
  \medskip

  \noindent
  \lbegin\\
  \SPC ${\mathcal F}_i \lla (\eb_i, f_i)$ for $i=1,\cdots,m$\\
  \SPC $B \lla \{{\mathcal F}_i  \mid  i=1,\cdots,m\}$\\
  \SPC $CP \lla \{\mbox{critical pair } [{\mathcal F}_i, {\mathcal F}_j] \mid  1\le i<j \le  m\}$\\
  \SPC  \lwhile $CP$ is not empty \ldo\\
  \SPC\SPC  $cp \lla $ select a critical pair from $CP$\\
  \SPC\SPC  $CP \lla CP \setminus \{cp\}$\\
  \SPC\SPC  \lif $cp$ meets {\bf neither} Syzygy Criterion {\bf nor} Rewritten
  Criterion,\\
  \SPC\SPC\SPC \lthen\\
  \SPC\SPC\SPC\SPC  ${\mathcal SP} \lla $ the S-polynomial of critical pair
  $cp$\\
  \SPC\SPC\SPC\SPC  ${\mathcal P} \lla $ the F5-reduction result of ${\mathcal SP}$ by $B$, i.e. ${\mathcal SP}\Longrightarrow_B^* {\mathcal P}$\\
  \SPC\SPC\SPC\SPC  \lif the polynomial of ${\mathcal P}$ is {\bf not} $0$, i.e. $\p({\mathcal P})\not= 0$,\\
  \SPC\SPC\SPC\SPC\SPC \lthen \\
  \SPC\SPC\SPC\SPC\SPC\SPC $CP \lla CP \cup \{\mbox{critical pair } [{\mathcal P}, {\mathcal Q}] \mid {\mathcal Q}\in B\}$\\
  \SPC\SPC\SPC\SPC \lendif\\
  \SPC\SPC\SPC\SPC  $B \lla B \cup \{{\mathcal P}\}$ \SPC $\#$ no matter whether $\p({\mathcal P})\not= 0$ or not\\
  \SPC\SPC \lendif\\
  \SPC \lendwhile\\
  \SPC \lreturn $\{\mbox{polynomial part of } {\mathcal Q} \mid {\mathcal Q}\in B\}$\\
  \lend
  \medskip
\end{algorithm}

Of course, some auxiliary data are also necessary to be recorded in the implementation. For example, we need to keep   the
generating order of the labeled polynomials. We can also save labeled polynomials as $(\lpp(\gb), g)$ instead of $(\gb, g)$
during the compuation, since only the information of $\lpp(\gb)$ is really used.

The strategy of selecting critical pairs is not specified in the F5B
algorithm, instead we simply use $$cp \lla \mbox{ select a critical
pair from } CP.$$ Different strategies of selecting critical pairs
will lead to different versions of the F5 algorithm, including the
original {\em Incremental} F5 algorithm in \citep{Fau02}, {\em
Non-Incremental} F5 algorithm
reported by Fau\`gere in recent conference (INSCRYPT 2008), {\em
Matrix} F5 algorithm mentioned in \citep{Bardet03}. All these
versions of the F5 algorithm are discussed in the next section.

Moveover, the original F5 algorithm differs from the F5B algorithm only by a strategy of selecting critical pairs as well. Let us
see more discussions in the following subsection.

\subsection{Equivalence of the F5 and F5B algorithms}
\label{subsec-equivalence}

The major difference between   F5   \citep{Fau02} and F5B   is the reduction process. In \citep{Fau02}, reduction of
S-polynomials is done by the {\em Reduction} function. So next, we will focus on showing that the {\em Reduction} function is
equivalent to a set of F5-reduction with an appropriate strategy of selecting critical pairs. Let us see the {\em TopReduction}
function first, which is a subfunction of the {\em Reduction} function (function \ref{fun-topreduce}).

\begin{function} \caption{\bf --- TopReduction} \label{fun-topreduce}
  \smallskip
  \Input{a labeled polynomial ${\mathcal F}$, a set of labeled polynomials $B$, a set of labeled polynomials $Todo$ which needs to be reduced further. }\\
  \Output{a $2$-tuple $(Done', Todo')$: both $Done'$ and $Todo'$ are sets of labeled polynomials,
  while $Done'$ contains the result of reduction and $Todo'$
  includes the labeled polynomials to be reduced further.}
  \medskip

  \noindent
  \lbegin\\
  \SPC ${\mathcal G} \lla $ a labeled polynomial in $B$ such
  that: \\
  \SPC\SPC\SPC\SPC (1) $\lpp({\mathcal G}) \mid \lpp({\mathcal F})$ and denote $v=\lm({\mathcal F})/\lm({\mathcal
  G})$, \\
  \SPC\SPC\SPC\SPC (2) $v{\mathcal G}$ is {\bf not} divisible by
  $B$, and \\
  \SPC\SPC\SPC\SPC (3) $v{\mathcal G}$ is {\bf not} rewritable by
  $B \cup \{ {\mathcal F} \} \cup Todo$. \\
  \SPC  \lif such ${\mathcal G}$ does not exist, \\
  \SPC\SPC \lthen \\
  \SPC\SPC\SPC \lreturn $(\{{\mathcal F}\}, \emptyset)$\\
  \SPC\SPC \lelse \\
  \SPC\SPC\SPC \lif signature $\s({\mathcal F}) \sus \s(v{\mathcal
  G})$,\\
  \SPC\SPC\SPC\SPC \lthen \\
  \SPC\SPC\SPC\SPC\SPC \lreturn $(\emptyset, \{{\mathcal F}-v{\mathcal
  G})\}$\\
  \SPC\SPC\SPC\SPC \lelse \\
  \SPC\SPC\SPC\SPC\SPC \lreturn $(\emptyset, \{{\mathcal F}, v{\mathcal
  G}-{\mathcal F})\}$\\
  \SPC\SPC\SPC \lendif\\
  \SPC \lendif\\
  \lend
  \medskip
\end{function}


Next is the main function of {\em Reduction} in \citep{Fau02}
(function \ref{fun-reduce}).

\begin{function} \caption{\bf --- Reduction} \label{fun-reduce}
  \smallskip
  \Input{a set of labeled polynomials $Todo$ which need to be reduced, a set of labeled polynomials $B$.}\\
  \Output{the reduction results of $Todo$ reducing by the set $B$.}
  \medskip

  \noindent
  \lbegin\\
  \SPC $Done \lla \emptyset$\\
  \SPC \lwhile set $Todo$ is not empty, \\
  \SPC\SPC \ldo \\
  \SPC\SPC\SPC ${\mathcal F} \lla $ the labeled polynomial with minimal signature in set $Todo$\\
  \SPC\SPC\SPC $Todo \lla Todo \setminus \{{\mathcal F}\}$,\\
  \SPC\SPC\SPC $(Done', Todo') \lla TopReduction({\mathcal F}, B,Todo)$\\
  \SPC\SPC\SPC $Done \lla Done \cup Done'$\\
  \SPC\SPC\SPC $Todo \lla Todo \cup Todo'$\\
  \SPC \lendwhile\\
  \SPC \lreturn $Done$\\
  \lend
  \medskip
\end{function}

We can also write  F5-reduction as a function which is similar to the {\em TopReduction} (function \ref{fun-f5reduce}).

\begin{function} \caption{\bf --- F5-Reduction} \label{fun-f5reduce}
  \smallskip
  \Input{a labeled polynomial ${\mathcal F}$, a set of labeled polynomisl $B$.}\\
  \Output{a $2$-tuple $(Done, Todo)$: both $Done$ and $Todo$ are sets of lebeled polynomials,
  while $Done$ contains the result of reduction and $Todo$
  includes the labeled polynomials to be reduced further.}
  \medskip

  \noindent
  \lbegin\\
  \SPC ${\mathcal G} \lla $ a labeled polynomial in $B$ such
  that: \\
  \SPC\SPC\SPC\SPC (1) $\lpp({\mathcal G}) \mid \lpp({\mathcal F})$ and denote $v=\lm({\mathcal F})/\lm({\mathcal
  G})$, \\
  \SPC\SPC\SPC\SPC (2) signature $\s({\mathcal F}) \sus \s(v{\mathcal
  G})$,\\
  \SPC\SPC\SPC\SPC (3) $v{\mathcal G}$ is {\bf not} divisible by
  $B$, and \\
  \SPC\SPC\SPC\SPC (4) $v{\mathcal G}$ is {\bf not} rewritable by $B$. \\
  \SPC  \lif such ${\mathcal G}$ does not exist, \\
  \SPC\SPC \lthen \\
  \SPC\SPC\SPC \lreturn $(\{{\mathcal F}\}, \emptyset)$\\
  \SPC\SPC \lelse \\
  \SPC\SPC\SPC \lreturn $(\emptyset, \{{\mathcal F}-v{\mathcal
  G}\}$\\
  \SPC \lendif\\
  \lend
  \medskip
\end{function}

Thus, the F5-reduction can be used to replace the {\em TopReduction} function perfectly in the {\em Reduction} function by simply
updating the following codes:
$$(Done', Todo') \lla \mbox{F5-Reduction}({\mathcal F}, B).$$

Next, let us compare  {\em TopReduction} and F5-reduction.  Let ${\mathcal G}\in B$, $v=\lm({\mathcal F})/\lm({\mathcal G})$, and
suppose  $v{\mathcal G}$ is neither  divisible    nor rewritable by $B$ or $R$.

(1) For the case $\s({\mathcal F}) \sus \s(v{\mathcal G})$, there is
no difference between the {\em TopReduction} and F5-reduction.

(2) For the case $\s({\mathcal F}) = \s(v{\mathcal G})$:

\begin{enumerate}

\item[$\bullet$] In {\em TopReduction},
labeled polynomial $v{\mathcal G}$ is rewritable by $B\cup \{{\mathcal F}\}$ . Thus the labeled polynomial ${\mathcal G}$ can
never be selected from $B$, and nothing is done about ${\mathcal G}$ in {\em TopReduction}.

\item[$\bullet$] In F5-reduction, no matter which labeled polynomial is used
to F5-reduce ${\mathcal F}$, the reduction result of ${\mathcal F}$
has the same signature with ${\mathcal F}$ and will add to the set
$B$. So when discussing the critical pair $[{\mathcal F}, {\mathcal
G}]$ in later loops, this critical pair meets the Rewritten
Criterion  and is hence rejected. Thus no new labeled polynomial is
generated.
\end{enumerate}

(3) For the case $\s({\mathcal F}) <  v\s({\mathcal G})$:
\begin{enumerate}

\item[$\bullet$] In {\em TopReduction}, the labeled polynomial
$v{\mathcal G}-{\mathcal F}$ is calculated immediately and added to set $Todo$. Both labeled polynomials ${\mathcal F}$ and
$v{\mathcal G}$ will be reduced in later loops of the {\em Reduction} function. Notice that the labeled polynomial ${\mathcal F}$
may be selected from set $Todo$ in later loops, but in that time, the labeled polynomial ${\mathcal G}$ is not qualified to
reduce ${\mathcal F}$, as $v{\mathcal G}-{\mathcal F}$ has been added to the set $Todo$ and hence $v{\mathcal G}$ is rewritable
by $B \cup \{{\mathcal F}\}\cup Todo$.

\item[$\bullet$] While in F5-reduction, we cannot use the labeled polynomial
${\mathcal G}$ to F5-reduce ${\mathcal F}$ immediately. However, in the F5B algorithm, the critical pair $[{\mathcal F},
{\mathcal G}]$ must be added to the set $CP$, since labeled polynomial ${\mathcal G}$ is in the set $B$. Moreover,
  the S-polynomial of the critical pair $[{\mathcal F}, {\mathcal G}]$ is exactly the
labeled polynomial $v{\mathcal G}-{\mathcal F}$.  The S-polynomial $\sp({\mathcal F}, {\mathcal G})$ which can also be computed
immediately in the  next step if we use an appropriate strategy which selects the critical pair $[{\mathcal F}, {\mathcal G}]$
first.
\end{enumerate}

The third case $\s({\mathcal F}) <  v\s({\mathcal G})$ is most
complicated, so we illustrate the last case by a simple example.

\begin{example}
Consider a system in $\Q[x, y, z]$ with the Graded Reverse Lex
Order ($x\succ y\succ z$):
$$\left\{\begin{array}{l} f_1=xz^2+y^2, \\
f_2=xy+xz, \\
f_3=yz+z.
\end{array}\right. $$
\end{example}

The corresponding labeled polynomials of $f_1$, $f_2$ and $f_3$
are ${\mathcal F}_1=(\eb_1, f_1)$, ${\mathcal F}_2=(\eb_2, f_2)$
and ${\mathcal F}_3=(\eb_3, f_3)$, respectively. The labeled
polynomial set is $B=\{{\mathcal F}_1, {\mathcal F}_2, {\mathcal
F}_3\}$. We need to consider the critical pairs $[{\mathcal F}_1,
{\mathcal F}_2]=(y, {\mathcal F}_1, z^2, {\mathcal F}_2)$,
$[{\mathcal F}_1, {\mathcal F}_3]=(y, {\mathcal F}_1, xz,
{\mathcal F}_3)$ and $[{\mathcal F}_2, {\mathcal F}_3]=(z,
{\mathcal F}_2, x, {\mathcal F}_3)$.

Since the degree of $\lcm(\lpp({\mathcal F}_2), \lpp({\mathcal
F}_3))=xyz$ is $3$, which is lower than the other $\lcm$'s, we
operate the critical pair $[{\mathcal F}_2, {\mathcal F}_3]$ first.
Now we can see the major difference between {\em TopReduction} and
F5-reduction, since there exists a labeled polynomial ${\mathcal
F}_1$ such that $\lpp({\mathcal F}_1)=xz^2 \mid
xz^2=\lpp({\mathcal P})$ but signature $\s({\mathcal P}) \prs
\s({\mathcal F}_1)$.
\begin{enumerate}

\item[$\bullet$] In {\em TopReduction}, the labeled polynomial
${\mathcal Q} = {\mathcal F}_1 - {\mathcal P}=(\eb_1-z\eb_2+x\eb_3, xz+y^2)$ is calculated. Next, both labeled polynomials
${\mathcal P}$ and ${\mathcal Q}$ add to the set $Todo$ for further reductions. In the next loop of the {\em Reduction} function,
the labeled polynomial ${\mathcal P}=(z\eb_2-x\eb_3, xz^2-xz)$ is irreducible, since the labeled polynomial ${\mathcal F}_1$ is
rewritable by  $R$ this time. The labeled polynomial ${\mathcal Q} =(\eb_1, xz+y^2)$ is irreducible as well. Then both ${\mathcal
P}$ and ${\mathcal Q}$ add to set $Done$ and create new critical pairs $\{[{\mathcal F}_1, {\mathcal P}], [{\mathcal P},
{\mathcal F}_2], [{\mathcal P}, {\mathcal F}_3], [{\mathcal F}_1, {\mathcal Q}], [{\mathcal Q}, {\mathcal F}_2], [{\mathcal Q},
{\mathcal F}_3]$, $[{\mathcal Q}, {\mathcal P}]\}$. Combined with previous un-operated critical pairs $\{[{\mathcal F}_1,
{\mathcal F}_2], [{\mathcal F}_1, {\mathcal F}_3]\}$, the F5 algorithm continues to operate on these critical pairs.

\item[$\bullet$] While in F5-reduction, the labeled polynomial
${\mathcal P}$ is not F5-reducible by set $B=\{{\mathcal F}_1,
{\mathcal F}_2, {\mathcal F}_3\}$. So the labeled polynomial
${\mathcal P}$ adds to set $B$ immediately, and creates new critical
pairs $\{[{\mathcal F}_1, {\mathcal P}], [{\mathcal P}, {\mathcal
F}_2]$, $[{\mathcal P}, {\mathcal F}_3]\}$. Notice that there are
still two un-operated critical pairs $\{[{\mathcal F}_1, {\mathcal
F}_2], [{\mathcal F}_1, {\mathcal F}_3]\}$. When using a strategy of
selecting critical pairs that selects the critical pair $[{\mathcal
F}_1, {\mathcal P}]$ first, then the S-polynomial $\sp({\mathcal
F}_1, {\mathcal P})={\mathcal F}_1 - {\mathcal P}$ is calculated,
and obtain ${\mathcal Q} = {\mathcal F}_1 - {\mathcal
P}=(\eb_1-z\eb_2+x\eb_3, xz+y^2)$, which is not F5-reducible by set
$\{{\mathcal F}_1, {\mathcal F}_2, {\mathcal F}_3, {\mathcal P}\}$.
Then the new critical pairs $\{[{\mathcal F}_1, {\mathcal Q}],
[{\mathcal Q}, {\mathcal F}_2], [{\mathcal Q}, {\mathcal F}_3],
[{\mathcal Q}, {\mathcal P}]\}$ are created as well.
\end{enumerate}

In both cases, the remaining  critical pairs are $\{[{\mathcal
P}, {\mathcal F}_2], [{\mathcal P}, {\mathcal F}_3], [{\mathcal
F}_1, {\mathcal Q}], [{\mathcal Q}, {\mathcal F}_2], [{\mathcal
Q}, {\mathcal F}_3]$, $[{\mathcal Q}, {\mathcal P}]$, $[{\mathcal
F}_1, {\mathcal F}_2]$, $[{\mathcal F}_1, {\mathcal F}_3]\}$,
since critical pair $[{\mathcal F}_1, {\mathcal P}]$ will be
rejected by Rewritten Criterion in {\em TopReduction}.

After all, $TopReduciton$ is equivalent to F5-reduction and hence
the F5 algorithm is equivalent to the F5B algorithm with an
appropriate strategy of selecting critical pairs. However, the F5B
algorithm is simpler and easy to understand and analyze. Moreover,
by using different strategies (of selecting critical pairs), the F5B
algorithm becomes the {\em Incremental} F5 algorithm, {\em
Non-incremental} F5 algorithm and {\em Matrix} F5 algorithm which
are introduced in the next section.

Since the F5B algorithm is equivalent to the F5 algorithm but much
simpler to analyze, we propose a new proof for the correctness of
the F5B algorithm in \citep{SunWang10}. This new proof does not
depend on the strategies of selecting critical pairs, so it also
proves the correctness for the other versions of the F5 algorithm.
From this new proof, we find that the key of the F5B or F5 algorithm
is the special reduction procedure which is a one-way reduction,
i.e. only labeled polynomials with lower signatures can be used to
reduce labeled polynomials with higher signatures. This fact will be
much clearer in the {\it Matrix} F5 algorithm introduced later.

\section{Some Versions of the F5 Algorithm} \label{sec-presentversion}

In this section, we discuss three versions of the F5 algorithm. They
are {\em Incremental} F5 algorithm, {\em Non-incremental} F5
algorithm and {\em Matrix} F5 algorithm.

\subsection{Incremental F5 algorithm}

{\em Incremental} F5 algorithm is the original F5 algorithm
presented in \citep{Fau02}. If the F5B algorithm uses an appropriate
strategy of selecting critical pairs, then the F5B algorithm becomes
the {\em Incremental} F5 algorithm.

To describe this strategy appropriately, we need some new
notations. For a labeled polynomial ${\mathcal F}=(\fb, f)$ whose
signature is $\s({\mathcal F})=\lpp(\fb)=x^\alpha \eb_i$, we
define the index of ${\mathcal F}$ to be $\ind({\mathcal F}):=i$.
Given a critical pair $[{\mathcal F}, {\mathcal G}]$, we also
define $\ind([{\mathcal F}, {\mathcal G}]):=\min\{\ind({\mathcal
F}), \ind({\mathcal G})\}$. Then the strategy for incremental F5B
algorithms can be described as follows:
$$j \lla \max\{\ind([{\mathcal F}, {\mathcal
G}])\mid [{\mathcal F}, {\mathcal G}]\in CP\},$$
$$cp \lla \mbox{ select a critical
pair from } CP \mbox{ with index } j.$$

Let us see an easy example. Let $\{f_1, f_2, f_3, f_4\}\subset
\K[X]$ be the initial polynomials. Then at the beginning, set $CP$
contains all the critical pairs: $[{\mathcal F}_1, {\mathcal F}_2]$,
$[{\mathcal F}_1, {\mathcal F}_3]$, $\cdots$, $[{\mathcal F}_3,
{\mathcal F}_4]$. By the above strategy, $j=3$ at the beginning. So
the critical pairs with index $1$ or $2$ (such as $[{\mathcal F}_2,
{\mathcal F}_4]$) cannot be selected unless all the critical pairs
with index $3$ have been operated. This means the critical pairs
with index $1$ or $2$ can only be operated when the \gr basis of
ideal $\langle f_3, f_4\rangle$ is obtained. Similarly, critical
pairs with index $1$ (such as $[{\mathcal F}_1, {\mathcal F}_3]$)
can be seleceted/operated only when the \gr basis of ideal $\langle
f_2, f_3, f_4\rangle$ is computed.

\subsection{Non-incremental F5 algorithm}

Faug\`ere also presents a {\em Non-incremental} F5 algorithm in a
recent
conference (INSCRPT 2008). 
For example, it can use a strategy of selecting critical pairs like:
$$d \lla \min\{\deg(\lcm(\lpp({\mathcal F}), \lpp({\mathcal
G})))\mid [{\mathcal F}, {\mathcal G}]\in CP\},$$
$$cp \lla \mbox{ select a critical
pair from } CP \mbox{ with degree } d.$$

In this case, for example, the critical pair $[{\mathcal F}_1,
{\mathcal F}_3]$ with index $1$ may be operated earlier (not
necessarily after the \gr basis of ideal $\langle f_2, \cdots,
f_m\rangle$ is obtained).

However, this {\em Non-incremental} F5 algorithm is not really
non-incremental. Since all the critical pairs should be operated
sooner or later, the output of algorithm still contains the \gr
bases of the ideal $\langle f_i, \cdots, f_m\rangle$ where $1<i<m$.
To transform the F5 algorithm to a real non-incremental algorithm,
we need to change the order of signatures. This is detailed in
\citep{Ars09, SunWang10}.

\subsection{Matrix F5 algorithm}

The most efficient version of the F5 algorithm is the {\em Matrix}
F5 algorithm, which is an F5 algorithm that utilizes the matrix
techniques introduced by the F4 algorithm when reducing
S-polynomials. The {\em Matrix} F5 algorithm improves the original
F5 algorithm in data structure, so the {\em Matrix} F5 algorithm
differs from the F5B algorithm only by a strategy of selecting
critical pairs as well. Let us see the following example from
\citep{Fau02}.

\begin{example}
Consider the homogeneous system in $\F_{23}[x, y, z]$ with the
Graded Reverse Lex Order ($x\succ y\succ z$) by the Matrix F5
algorithm \footnote{Here we set parameter $b=0$ directly. }
$$\left\{\begin{array}{l} f_3=x^2+18xy+19y^2+8xz+5yz+7z^2,  \\
f_2=3x^2+7xy+22xz+11yz+22z^2+8y^2, \\
f_1=6x^2+12xy+4y^2+14xz+9yz+7z^2.
\end{array}\right. $$
\end{example}

In order to compute the \gr basis of $\langle f_1, f_2, f_3\rangle$, we set
${\mathcal F}_1=(\eb_1, f_1)$, ${\mathcal F}_2=(\eb_2, f_2)$ and
${\mathcal F}_3=(\eb_3, f_3)$. Next consider the critical pairs
$[{\mathcal F}_1, {\mathcal F}_2]=(1, {\mathcal F}_1, 1, {\mathcal
F}_2)$, $[{\mathcal F}_1, {\mathcal F}_3]=(1, {\mathcal F}_1, 1,
{\mathcal F}_3)$ and $[{\mathcal F}_2, {\mathcal F}_3]=(1,
{\mathcal F}_2, 1, {\mathcal F}_3)$. Like the F4 algorithm, we use
the part $1\times f_1, 1 \times f_2, 1 \times f_3$ to build the
matrix of degree $2$ in order to reduce the S-polynomials
generated from these three critical pairs together:

\[
\begin{array}{cc}
 & \begin{array}{cccccc}
x^2 & xy & y^2 & xz & yz & z^2
 \end{array}\\
\begin{array}{r}
f_3 \\ A_2 = \ f_2 \\ f_1
\end{array} &
\left(
\begin{array}{cccccc}
1 & 18 & 19 & 8 & 5 & 7\\
3 & 7 & 8 & 22 & 11 & 22\\
6 & 12 & 4 & 14 & 9 & 7
\end{array}
\right)
\end{array}
\]
and after triangulation of the matrix $A_2$:
\[
\begin{array}{cc}
 & \begin{array}{cccccc}
x^2 & xy & y^2 & xz & yz & z^2
 \end{array}\\
\begin{array}{r}
f_3 \\ B_2 = \ \underline{f_2} \\ f_1
\end{array} &
\left(
\begin{array}{cccccc}
1 & \underline{18} & 19 & 8 & 5 & 7\\
0 & 1 & 3 & 2 & 4 & -1\\
0 & \underline{0} & 1 & -11 & -3 & -5
\end{array}
\right)
\end{array}
\]
and two ``new" polynomials appear: $f_4=xy+4yz+2xz+3y^2-z^2$
(${\mathcal F}_4=(-\eb_2+3\eb_3, f_4)$ and $\s({\mathcal
F}_4)=\eb_2$) and $f_5=y^2-11xz-3yz-5z^2$ (${\mathcal
F}_5=(-4\eb_1+16\eb_2-\eb_3, f_5)$ and $\s({\mathcal F}_5)=\eb_1$),
which are the reduction results of the S-polynomials $\sp({\mathcal
F}_1, {\mathcal F}_2)$, $\sp({\mathcal F}_1, {\mathcal F}_3)$ and
$\sp({\mathcal F}_2, {\mathcal F}_3)$.

For simplifying the statement, we will use the signatures of polynomials to replace the signatures of the
 corresponding labeled polynomials without  confusions in the following of this example.

 Notice that the signature of
the polynomial $f_4$ is $\eb_2$, which corresponds to the label on
the left of that row (underlined $\underline{f_2}$ in the matrix
$B_2$).

Also we remark that the underlined $\underline{18}$ is not reduced
by $f_4$ since the signature of $f_3$ is $\eb_3$ which is smaller
than $\eb_2$ (the signature of $f_4$). While the underlined
$\underline{0}$ is reduced, since $\eb_1 \succ \eb_2$. This shows
that the reduction procedure in the F5 algorithm is a one-way
reduction.

The next step is to consider the newly generated critical pairs:
$[{\mathcal F}_1, {\mathcal F}_4]=(y, {\mathcal F}_1, x, {\mathcal
F}_4)$, $[{\mathcal F}_4, {\mathcal F}_2]=(x, {\mathcal F}_4, y,
{\mathcal F}_2)$, $[{\mathcal F}_4, {\mathcal F}_3]=(x, {\mathcal
F}_4, y, {\mathcal F}_3)$, $[{\mathcal F}_5, {\mathcal F}_4]=(x,
{\mathcal F}_5, y, {\mathcal F}_4)$, $[{\mathcal F}_5, {\mathcal
F}_1]=(x^2, {\mathcal F}_5$, $y^2, {\mathcal F}_1)$, $[{\mathcal
F}_5, {\mathcal F}_2]=(x^2, {\mathcal F}_5, y^2, {\mathcal F}_2)$
and $[{\mathcal F}_5, {\mathcal F}_3]=(x^2, {\mathcal F}_5, y^2,
{\mathcal F}_3)$. We select these pairs by degree and build the
matrix $A_3$ of degree $3$ in order to operate the following
critical pairs
$$[{\mathcal F}_1, {\mathcal F}_4]=(y, {\mathcal F}_1, x, {\mathcal F}_4), [{\mathcal
F}_4, {\mathcal F}_2]=(x, {\mathcal F}_4, y, {\mathcal F}_2),
[{\mathcal F}_4, {\mathcal F}_3]=(x, {\mathcal F}_4, y, {\mathcal
F}_3),$$ $$[{\mathcal F}_5, {\mathcal F}_4]=(x, {\mathcal F}_5, y,
{\mathcal F}_4)$$ together. We only need to consider the parts
$y\times f_3, y\times f_4, x\times f_4, x\times f_5$, since the
parts $y\times f_2, y\times f_1$ are rewritable by ${\mathcal
F}_4$ and ${\mathcal F}_5$ respectively.

Like the F4 algorithm, the parts $y\times f_3, y\times f_4, x\times
f_4, x\times f_5$ are the rows to be reduced in the matrix, and we
also need to select rows that are use to reduce these rows. Since
power products $y^3, x^2z, xyz, y^2z$ appear in the parts $yf_3,
yf_4, xf_4, xf_5$, we should add parts $yf_5, xf_3, zf_4, zf_5$ to
matrix $A_3$ in order to eliminate these power products.

Now we have the matrix $A_3$ of degree $3$. This matrix is ordered
by the signatures of each row, which are listed in the round
brackets:
\[
\begin{array}{cc} & \begin{array}{cccccccccc}
x^2y & xy^2 & y^3 & x^2z & xyz & y^2z & xz^2 & yz^2 & z^3
\end{array}\\
\begin{array}{r}
zf_3\ (z\eb_3)\\ yf_3\ (y\eb_3)\\ zf_4\ (z\eb_2)\\ A_3=\ yf_4\
(y\eb_2)\\ xf_4\ (x\eb_2)\\ zf_5\ (z\eb_1)
\\ yf_5\ (y\eb_1)\\ xf_5\ (x\eb_1)
\end{array} &
\left(
\begin{array}{cccccccccc}
0 & 0 & 0 & 1 & 18 & 19 & 8 & 5 & 7\\
1 & 18 & 19 & 0 & 8 & 5 & 0 & 7 & 0\\
0 & 0 & 0 & 0 & 1 & 3 & 2 & 4 & 22\\
0 & 1 & 3 & 0 & 2 & 4 & 0 & 22 & 0\\
1 & 3 & 0 & 2 & 4 & 0 & 22 & 0 & 0\\
0 & 0 & 0 & 0 & 0 & 1 & 12 & 20 & 18\\
0 & 0 & 1 & 0 & 12 & 20 & 0 & 18 & 0\\
0 & 1 & 0 & 12 & 20 & 0 & 18 & 0 & 0
\end{array}
\right)
\end{array}
\]
and after triangulation (ordered by the leading power products of
each row):
\[
\begin{array}{cc} & \begin{array}{cccccccccc}
x^2y & xy^2 & y^3 & x^2z & xyz & y^2z & xz^2 & yz^2 & z^3
\end{array}\\
\begin{array}{r}
yf_3\ (y\eb_3)\\ yf_4\ (y\eb_2)\\ \underline{xf_4}\ (x\eb_2)\\
B_3=\ zf_3\ (z\eb_3)\\ zf_4\ (z\eb_2)\\ zf_5\ (z\eb_1)
\\ \underline{yf_5}\ (y\eb_1)\\ \underline{xf_5}\ (x\eb_1)
\end{array} &
\left(
\begin{array}{cccccccccc}
1 & 18 & 19 & 0 & {8} & 5 & 0 & 7 & 0\\
0 & 1 & 3 & 0 & {2} & 4 & 0 & 22 & 0\\
0 & 0 & \textbf{1} & \underline{0} & \underline{0} & \underline{8} & \underline{1} & \underline{18} & 15\\
0 & 0 & 0 & 1 & 18 & 19 & 8 & 5 & 7\\
0 & 0 & 0 & 0 & 1 & 3 & 2 & 4 & 22\\
0 & 0 & 0 & 0 & 0 & 1 & 12 & 20 & 18\\
0 & 0 & 0 & 0 & 0 & 0 & \textbf{1} & 11 & 13\\
0 & 0 & 0 & 0 & 0 & 0 & 0 & \textbf{1} & 18
\end{array} \right)
\end{array}
\]
and the polynomials $f_6=y^3+8y^z+xz^2+18yz^2+15z^3$ (${\mathcal
F}_6=((15x+18y+12z)\eb_2+(x+7y+17z)\eb_3, f_6)$ and $\s({\mathcal
F}_6)=x\eb_2$), $f_7=xz^2+11yz^2+13z^3$ (${\mathcal
F}_7=((18y+18z)\eb_1+(10x+9y+20z)\eb_2+(16x+13y+13z)\eb_3, f_7)$ and
$\s({\mathcal F}_7)=y\eb_1$) and $f_8=yz^2+18z^3$ (${\mathcal
F}_8=((11x+11y-3z)\eb_1+(21y+9z)\eb_2+(3x+7y-3z)\eb_3, f_8)$ and
$\s({\mathcal F}_8)=x\eb_1$) are the reduction results of the
S-polynomials of degree 3. Notice that although
$\lpp(f_6)=y^3=y\lpp(f_5)$, labeled polynomial ${\mathcal F}_6$ is
not F5-reducible by ${\mathcal F}_5$. Thus $f_6$ is still a ``new"
polynomial.

Now the Rewritten Criterion is much clearer. When building the
matrix $A_3$, we list the signatures of each row in round brackets.
Labeled polynomials with the same signatures will play the same role
in the matrix, so among the labeled polynomials with the same
signatures, it suffices to deal with the latest results (that is why
we care about the creating order of labeled
polynomials).\footnote{If we add the row $xf_3\ (x\eb_3)$ to matrix
$A_3$, replacing the polynomials $zf_4, yf_4, xf_4, zf_5, yf_5,
zf_5$ by polynomials $zf_2$, $yf_2$, $xf_2$, $zf_1$, $yf_1$, $xf_1$
will lead to the same triangular form $B_3$.}

Also the one-way reduction is evident in the matrix $B_3$. Let us
see the row $\underline{xf_4}\ (x\eb_2)$. The underlined
$\underline{0}$, $\underline{0}$ are reduced by rows $zf_3\
(z\eb_3)$ and $zf_4\ (z\eb_2)$ respectively, while underlined
$\underline{8}$, $\underline{1}$, $\underline{18}$ are not
eliminated by rows $zf_5\ (z\eb_1)$, $\underline{yf_5}\ (y\eb_1)$
and $\underline{xf_5}\ (x\eb_1)$. The reason lies in the one-way
reduction. More specifically, the signatures of rows $zf_3\
(z\eb_3)$ and $zf_4\ (z\eb_2)$ are $z\eb_3$ and $z\eb_2$, both of
which are smaller than the signature $x\eb_2$ of row
$\underline{xf_4}\ (x\eb_2)$. Thus, the rows  $zf_3\ (z\eb_3)$ and
$zf_4\ (z\eb_2)$ are able to reduce row $\underline{xf_4}\
(x\eb_2)$. However, we have signatures $z\eb_1, y\eb_1, x\eb_1 \succ
x\eb_2$, so the rows $zf_5\ (z\eb_1)$, $\underline{yf_5}\ (y\eb_1)$
and $\underline{xf_5}\ (x\eb_1)$ are not qualified to reduce the row
$\underline{xf_4}\ (x\eb_2)$. Remark that, since only the rows
$\underline{xf_4}\ (x\eb_2)$, $\underline{yf_5}\ (y\eb_1)$ and
$\underline{xf_5}\ (x\eb_1)$ are worth saving, the others rows are
not fully reduced in matrix $B_3$.

However, we must realize that, although the two new criteria of the
F5 algorithm could reject almost all useless computations, the
one-way reduction results in a poorer efficiency of eliminating
matrix than the F4 algorithm. So it is really difficult to tell
which of the F4 and F5 algorithms is faster the other, especially in
large examples.

\section{Conclusions}

In this paper, we rewrite  F5 as F5B in Buchberger's style. We show that the F5B algorithm is equivalent to the original F5
algorithm. It is very easy to understand and implement F5B.
 The key of the F5B algorithm is the F5-reduction, which is a one-way reduction according to the signatures and their generating orders.
  F5B  is also a useful tool to analyze  F5 and F5-like algorithms.  Although the F5 algorithm
has good theoretical results for avoiding useless computations,  the one-way reduction slows down the efficiency of algorithm
more or less. It is desirable to have a more careful analysis of this issue, and we hope to work on it in the future.

\section{Acknowledgements}

We would like to thank S.H. Gao and X.S. Gao for   helpful discussions and suggestions.

\end{document}